\documentclass[twocolumn,showpacs,prd,floatfix,axodraw]{revtex4}
\usepackage{mathrsfs}
\usepackage{graphics}
\usepackage{graphicx,booktabs,bm}
\usepackage{overpic}
\usepackage{color}
\usepackage{amssymb}

\usepackage{mathrsfs,bm,amsmath,amssymb}
\usepackage{longtable,lscape}
\usepackage{txfonts}
\usepackage{amssymb}
\usepackage{indentfirst}
\usepackage{graphicx,,booktabs}
\usepackage{multirow}
\usepackage{color}
\usepackage{amssymb}

\definecolor{cover}{rgb}{0.77,0.87,0.88}
\definecolor{blueone}{rgb}{0.1,0.1,.7}
\definecolor{citec}{rgb}{0.14,0.47,0.09}
\definecolor{two}{rgb}{0.0,0.5,0.}
\definecolor{three}{rgb}{.5,.1,0.15}
\usepackage[bookmarks=true,bookmarksopen=false,plainpages=false,breaklinks=true,
   bookmarksnumbered=true,hypertexnames=false,
   filecolor=blue,urlcolor=three,menucolor=three,
   linkcolor=three,citecolor=blueone, colorlinks,
   anchorcolor=blue,runcolor=pink,frenchlinks=red
   pdfstartview=FitH,pdftitle=title,%
   pdfauthor=author]{hyperref}

\begin{document}
\title{{{Production of the $T^{+}_{cc}$ state in the $\gamma{}p\to{}D^{+}\bar{T}^{-}_{cc}\Lambda_c^{+}$ reaction}}}

\author{Yin Huang$^1$}
\author{Hong Qiang Zhu$^{2}$}
\email{20132013@cqnu.edu.cn}
\author{Li-Sheng Geng$^{3}$}
\email{lisheng.geng@buaa.edu.cn}
\author{Rong Wang$^{4,5}$}

\affiliation{$^1$School of Physical Science and Technology, Southwest Jiaotong University, Chengdu 610031,China\\
$^2$College of Physics and Electronic Engineering, Chongqing Normal University, Chongqing 401331,China\\
$^3$School of Physics and Nuclear Energy Engineering, Beihang University, Beijing 100191, China\\
$^4$Institute of Modern Physics, Chinese Academy of Sciences, Lanzhou 730000, China\\
$^5$University of Chinese Academy of Sciences, Beijing 100049, China}

\date{\today}
\begin{abstract}
Stimulated by the recent LHCb observation of a new exotic charged structure $T^{+}_{cc}$, we propose to use the central diffractive mechanism existing in the $\gamma{}p\to{}D^{+}\bar{T}^{-}_{cc}\Lambda_c^{+}$ ($\bar{T}_{cc}$ is antiparticle of $T^{+}_{cc}$)
reaction to produce $T^{+}_{cc}$.  Our theoretical approach is based on the chiral unitary theory where the $T^{+}_{cc}$
resonance is dynamically generated.  With the coupling constant of the $T^{+}_{cc}$ to $DD^{*}$ channel obtained from chiral unitary theory,
the total cross sections of the $\gamma{}p\to{}D^{+}\bar{T}^{-}_{cc}\Lambda_c^{+}$ reaction are evaluated.  Our study
indicates that the cross section for $\gamma{}p\to{}D^{+}\bar{T}^{-}_{cc}\Lambda_c^{+}$ reaction are of the order of 1.0 pb, which is accessible  at the proposed EicC~\cite{Anderle:2021wcy} and US-EIC~\cite{Accardi:2012qut} due to the higher luminosity.  If measured in
future experiments, the predicted total cross sections can be used to test the (molecular) nature of the $T^{+}_{cc}$.
\end{abstract}

%\pacs{13.60.Le, 12.39.Mk,13.25.Jx}

\maketitle
\section{INTRODUCTION}
Exotic hadrons~\cite{Zyla:2020zbs} has been a focus of theoretical and experimental interest since they have an
internal structure more complex than the simple $\bar{q}q$ configuration for mesons or $qqq$ configuration for baryons in the
traditional  constituent quark models~\cite{Gell-Mann:1964ewy}.  In particular, understanding their
production mechanism and using it as a probe to the structure of the hadrons are among the most active research fields
in particle and nuclear physics~\cite{Brambilla:2010cs}.  The challenge is to understand the nonperturbative transition from
high energy  $e^+e^-$, photon-hadron, and hadron-hadron collisions to physical exotic states.  In this work, we report a
central diffractive contribution existing in the $\gamma{}p\to{}D^{+}\bar{T}^{-}_{cc}\Lambda_c^{+}$ reaction to understand the
newly observed doubly charmed meson $T^+_{cc}$.

The charmed meson $T^+_{cc}$ with $I(J^{P})=0(1^{+})$ was first observed by the LHCb Collaboration in the $D^{0}D^{0}\pi^{+}$ invariant mass
spectrum~\cite{F.Muheim2021,I.Polyakov2021,LiupanAn2021}.  Its mass and width are measured to be
\begin{align}
m&=3875.09~{\rm MeV}+\delta{}m,\nonumber\\
\Gamma&=410\pm{}165\pm 43^{+18.0}_{-38}~{\rm KeV},
\end{align}
respectively.  From the $D^{0}D^{0}\pi^{+}$ decay mode, the new structure $T^+_{cc}$ contains at least four valence quarks.
Because the quark components of $D^{0}$ and $\pi^{+}$ mesons are $c\bar{u}$ and $u\bar{d}$, respectively,  the $T^+_{cc}$ is
another new candidates of tetraquark state with double-charm quarks following the previous observation of the doubly charmed
baryon state $\Xi_{cc}^{++}(3621)$~\cite{LHCb:2017iph}.

In fact, there were already a few theoretical studies on the existence of such a state before discovery.  A loosely $D^{(*)}D^{(*)}$
bound state with a binding energy of $0.47\sim42.82$ MeV was predicted by taking into account the coupled channel effect, and the isospin
(spin-parity) of the $T^{+}_{cc}$ state was suggested to be $I(J^{P})=0(1^{+})$~\cite{Li:2012ss}.   The mass spectra of doubly-charmed
meson state $T^{+}_{cc}$ around 3873 MeV was investigated in the one-boson-exchange model~\cite{Liu:2019stu} that can be
associated with the LHCb observation~\cite{F.Muheim2021,I.Polyakov2021,LiupanAn2021}. In Refs.~\cite{Mehen:2017nrh,Eichten:2017ffp},
a doubly charmed compact tetraquark state above the $DD^{*}$ mass threshold was predicted by considering the heavy quark symmetry.
The axial-vector tetraquark state of $cc\bar{u}\bar{d}$  was studied using the QCD sum rule method~\cite{Navarra:2007yw}.
Exotic mesons $J^P=0^{\pm}$ and $1^{\pm}$ with a general content $cc\bar{q}\bar{q}^{'}$ in  the same approach were
investigated in Ref.~\cite{Du:2012wp}.  The axial-vector state $cc\bar{u}\bar{d}$  was modeled as a hadronic molecule of $D^{0}D^{*+}$~\cite{Dias:2011mi}.

Following the discovery of the $T^{+}_{cc}$, several theoretical studies have been performed.  The mass and current coupling of the newly
observed doubly charmed four-quark state $T^{+}_{cc}$ is calculated in the QCD two-point sum rule method with the conclusion
that the $T^{+}_{cc}$ can be assigned as an axial-vector tetraquark~\cite{Agaev:2021vur}. Since the mass of  $T^{+}_{cc}$
is just 273 KeV below the $D^{*+}D^0$ threshold, it can naturally understood as a $DD^*$ molecule~\cite{Meng:2021jnw,Yan:2021wdl,Dong:2021bvy,Feijoo:2021ppq,Chen:2021vhg,Wu:2021kbu,Ling:2021bir}.
From  Refs.~\cite{Agaev:2021vur,Meng:2021jnw,Yan:2021wdl,Dong:2021bvy,Feijoo:2021ppq,Chen:2021vhg,Wu:2021kbu,Ling:2021bir},
it seems that the $T^{+}_{cc}$  can not  only be explained as a compact tetraquark state but also  as a molecular state.
Moreover, the parameters of the $T^{+}_{cc}$ state obtained from fits to the inclusive cross section~\cite{F.Muheim2021,I.Polyakov2021,LiupanAn2021}
are poorly understood theoretically, precise information on the production mechanism of the $T^{+}_{cc}$ is crucial in better understanding the $T^{+}_{cc}$.

In this work, we report on a theoretical study of  $\bar{T}^{-}_{cc}$ in $\gamma{}p\to{}D^{+}\bar{T}^{-}_{cc}\Lambda_c^{+}$ reaction
employing the central diffractive mechanism, which has been widely employed to
investigate the production of hadrons in $pp$ collisions~\cite{Lebiedowicz:2011tp,Lebiedowicz:2009vt,Lebiedowicz:2009pj,Kaidalov:1979jz}.
However, it was not studied in too much detail in photon-hadron reactions either experimentally or theoretically.
This is because at not too high energies the hadrons in the final state contribute negligibly  to central
diffractive productions compared with other processes~\cite{Kaidalov:1979jz}.  High energy photon beams are available at Electron-Ion Collider
in China (EicC)~\cite{Anderle:2021wcy} or US(US-EIC)~\cite{Accardi:2012qut},  which provide another alternative to study the
$\gamma{}p\to{}D^{+}\bar{T}^{-}_{cc}\Lambda_c^{+}$ reaction by considering the central diffractive mechanism.
The contributions for the $\gamma{}p\to{}D^{+}\bar{T}^{-}_{cc}\Lambda_c^{+}$ reaction from other channels, such as $s-$ and $u-$ channels,
are ignored because the $s-$ and $u-$ channels, which involve the creation of two additional $\bar{c}c$ quark pairs in the photon
induced production, are usually strongly suppressed.  Thus, the central diffractive mechanism provides the dominant contribution in the
$\gamma{}p\to{}D^{+}\bar{T}^{-}_{cc}\Lambda_c^{+}$ reaction and is a good way to test the nature of the $T^{+}_{cc}$.

This paper is organized as follows. In Sec.~\ref{Sec: formulism}, we
present the theoretical formalism. In Sec.~\ref{Sec: results}, the numerical
result of the $\bar{T}^{-}_{cc}$ in $\gamma{}p\to{}D^{+}\bar{T}^{-}_{cc}\Lambda_c^{+}$
will be given, followed by discussions and conclusions in the last section.

\section{THEORETICAL FORMALISM}\label{Sec: formulism}
\begin{figure}[htbp]
\begin{center}
\includegraphics[scale=0.5]{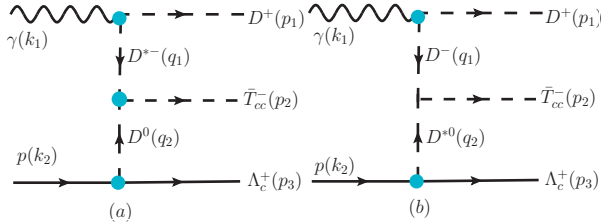}
\caption{ Central diffractive mechanism responsible for the production of $\bar{T}^{-}_{cc}$ in the $\gamma{}p$ collision where $\bar{T}^{-}_{cc}$
is treated as a $D^{*-}\bar{D}^{0}$ hadronic molecule. We also show the definition of the kinematics
$(p_1, p_2, p_3, k_1,k_2,q_1,q_2)$. The $\bar{T}^{-}_{cc}$ represents the antiparticle of $T^{+}_{cc}$.}\label{mku-coupling}
\end{center}
\end{figure}

First we explain the central diffractive mechanism responsible for the
$\gamma{}p\to{}D^{+}\bar{T}^{-}_{cc}\Lambda_c^{+}$ reaction.  Similar to the central
diffractive mechanism in $pp$ collisions,  at higher energies free $\bar{D}^{*}\bar{D}$
production is dominant, while at lower energies and the $\bar{D}^{*}$ and $\bar{D}$ interacts strongly and produce $\bar{T}^{-}_{cc}$ .  The tree-level Feynman diagrams are depicted
in Fig.~\ref{mku-coupling}.

To compute the diagrams shown in Fig.~\ref{mku-coupling}, we need the effective Lagrangian densities for the relevant interaction
vertice.  As mentioned in Refs.~\cite{Meng:2021jnw,Yan:2021wdl,Dong:2021bvy,Feijoo:2021ppq,Chen:2021vhg,Wu:2021kbu,Ling:2021bir}, the $T^{+}_{cc}$ resonances
can be identified as an $S$-wave $D^{*}D$ molecule. Because the lowest angular momentum gives the dominant contribution in the near-threshold region,
a constant amplitude can be used for the $S$-wave vertex of the $T_{cc}D^{*}D$.  Thus, the Lagrangian densities for the
$S$-wave coupling of $T^{+}_{cc}$ with its components can be writen down as~\cite{Guo:2014taa,Guo:2017jvc}
\begin{align}
{\cal{L}}_{T_{cc}D^{*}D}&=g_{T_{cc}}T_{cc}^{\mu\dagger}D^{*}_{\mu}D,\label{eq1}
\end{align}
where $g_{T_{cc}}$ is the coupling constant.  Using exactly the same strategy as Ref.~\cite{Feijoo:2021ppq}
a state barely bound that can be associated with currently the $T_{cc}$~\cite{F.Muheim2021,I.Polyakov2021,LiupanAn2021}
is found.  The couplings of the bound state to the coupled channels $D^{*+}D^{0}$ (channel 1) and $D^{*0}D^{+}$ (channel 2)
can be obtained from the residue of the scattering amplitude at the pole
position $z_R$, which reads
\begin{align}
T_{ij}=\frac{g_{ii}g_{jj}}{\sqrt{s}-Z_R},
\end{align}
where $g_{ii}$ is the coupling of the state to the $i-$th channel.  The coupling constants are found to be
\begin{align}
g_{T_{cc}D^{*+}D^{0}}=3.67~{\rm GeV}, g_{T_{cc}D^{*0}D^{+}}=-3.92~{\rm GeV};
\end{align}
and, as we can see, they are basically opposite to each other indicating that the $T^{+}_{cc}$ is a quite good $I=0$ bound state.
Detailed calculations and discussions can be found in Ref~\cite{Feijoo:2021ppq}.

In addition to the vertices described by Eq.~(\ref{eq1}), the following effective Lagrangians~\cite{Du:2020bqj,Huang:2018wgr} are needed to evaluate
 the Feynman diagram shown in Fig.~\ref{mku-coupling}.  Because we study the production rate of the
$T^{+}_{cc}$ in the near-threshold region, it is sufficient to consider effective Lagrangians that have the smallest number
of derivatives, which are given as follows~\cite{Du:2020bqj,Huang:2018wgr},
\begin{align}
{\cal{L}}_{\Lambda_cND^{*}}&=g_{D^{*}N\Lambda_c}\bar{\Lambda}_c\gamma^{\mu}ND^{*}_{\mu}+H.c.,\\
{\cal{L}}_{\Lambda_cND}&=ig_{DN\Lambda_c}\bar{\Lambda}_c\gamma_5ND+H.c.,\\
{\cal{L}}_{\gamma{}DD^{*}}&=g_{\gamma{}DD^{*}}\epsilon_{\mu\nu\alpha\beta}(\partial^{\mu}{\cal{A}}^{\nu})(\partial^{\alpha}D^{*\beta})D+H.c.,\\
{\cal{L}}_{\gamma{}DD}&=ie{\cal{A}}_{\mu}(D^{+}\partial^{\mu}D^{-}-\partial^{\mu}D^{+}D^{-})
\end{align}
where ${\cal{A}}^{\mu}$, $D$, $D^{*\mu}$, $N$ and $\Lambda_c$ are the photon, $D-$meson, $D^{*}-$meson, nucleon and $\Lambda_c^{+}$ baryon
fields, respectively. $e=\sqrt{4\pi\alpha}$ with $\alpha$ being the fine-structure constant.
Coupling constants $g_{DN\Lambda_c}=-13.98$ and $g_{D^{*}N\Lambda_c}=-5.20$ are obtained from the light-cone sum rules~\cite{Khodjamirian:2011sp,Khodjamirian:2011jp}.
The coupling constant $g_{D^{*}D^{}\gamma}$ is determined by the radiative decay widths of $D^{*}$,
\begin{align}
\Gamma_{D^{*\pm}\to{}D^{\pm}\gamma}=\frac{g^2_{D^{*}D^{}\gamma}(m^2_{D^{*}}-m^2_{D})^2}{48\pi{}m^2_{D^{*}}}|\vec{p}_{D}^{c.m}|.
\end{align}
where $\vec{p}_{D}^{c.m}$ is the three-vector momentum of the $D$ in the $D^{*}$ meson rest frame.  With $m_{D^{*}}=2.01$ GeV,
$m_{D}=1.87$ GeV and $\Gamma_{D^{*\pm}\to{}D^{\pm}\gamma}=0.979-1.704$ KeV~\cite{Zyla:2020zbs}, one obtains $g_{D^{*}D^{}\gamma}=0.173-0.228$ GeV$^{-1}$.
According to the lattice QCD and QCD sum rule calculations~\cite{Becirevic:2009xp, Zhu:1996qy}, the minus sign
for $g_{D^{*}D\gamma}$ is adopted.

In evaluating the scattering amplitudes of the $\gamma{}p\to{}D^{+}\bar{T}^{-}_{cc}\Lambda_c^{+}$ reaction, we need to include form factors because hadrons are not pointlike particles.  For the exchanged $D$ and $D^{*}$ mesons, we
apply a widely used pole form factor, which is
\begin{align}
{\cal{F}}_{i}=\frac{\Lambda_i^2-m_i^2}{\Lambda_i^2-q_i^2}~~~~~~~ i=D,D^{*},
\end{align}
where  $q_i$ and $m_i$ are the four-momentum and the mass of the exchanged $D^{(*)}$ meson, respectively.
The $\Lambda_i$ is the hard cutoff, and it can be directly related to the hadron size.  Empirically, the cutoff parameter
$\Lambda_i$ should be at least a few hundred MeV larger than the mass of the exchanged hadron.  Hence,
the $\Lambda_i=m_i+\alpha{}\Lambda_{QCD}$ and the QCD energy scale $\Lambda_{QCD}=220$ MeV is adopted in this work.
The $\alpha$ reflects the nonperturbative property of QCD at the low-energy scale and it can only be determined
from experimental data. In the following, it will be taken as a parameter and discussed later

The propagator for the $D$ meson is written as
\begin{align}
G_{D}(q)=\frac{i}{q^2-m^2_{D}}.
\end{align}
For the $D^{*}$ exchange, we take the propagator as
\begin{align}
G^{\mu\nu}_{D^{*}}(q)=\frac{i(-g^{\mu\nu}+q^{\mu}q^{\nu}/m^2_{D^{*}})}{q^2-m^2_{D^{*}}},
\end{align}
where $\mu$ and $\nu$ are the polarization indices of $D^{*}$.

With all these ingredients, the invariant scattering amplitude of the $\gamma{}p\to{}D^{+}\bar{T}^{-}_{cc}\Lambda_c^{+}$ reaction as
shown in Fig.~\ref{mku-coupling} can be constructed as
\begin{align}
-i{\cal{M}}_{j}=\bar{u}(p_3,\lambda_{\Lambda_c^{+}}){\cal{W}}_{j}^{\mu\nu}u(k_2,\lambda_p)\epsilon_{\nu}(k_1,\lambda_{\gamma})\epsilon^{*}_{\mu}(p_2,\lambda_{T^{-}_{cc}}),
\end{align}
where $j$ denotes diagrams $a$, $b$ of Fig.~1 or a contact term  contribution that will be explained below, while $u$
and $\epsilon$ are the Dirac spinor and  polarization vector, respectively. $\lambda_{\Lambda_c^{+}}$,$\lambda_p$,$\lambda_{T^{-}_{cc}}$ and $\lambda_{\gamma}$
are the helicities for the $\Lambda_c^{+}$, the proton, the $\bar{T}^{-}_{cc}$ and the photon, respectively.

Then the reduced amplitudes ${\cal{W}}_{j}^{\mu\nu}$ read as
\begin{align}
{\cal{W}}^{\mu\nu}_{a}&=g_a\gamma_5\epsilon_{\alpha\nu\beta\rho}k_{1}^{\alpha}q^{\beta}_{1}\frac{-g^{\mu\rho}+q_1^{\mu}q_1^{\rho}/m^2_{D^{*-}}}{q^2_1-m^2_{D^{*-}}+im_{D^{*-}}\Gamma_{D^{*-}}}\frac{{\cal{F}}_{a}}{q_2^2-m^2_{\bar{D}^{0}}},\\
{\cal{W}}^{\mu\nu}_{b}&=g_b\gamma_{\rho}(q_1^{\nu}-p^{\nu}_{1})\frac{-g^{\mu\rho}+q_2^{\mu}q_2^{\rho}/m^2_{\bar{D}^{*0}}}{q^2_2-m^2_{\bar{D}^{*0}}+im_{\bar{D}^{*0}}\Gamma_{\bar{D}^{*0}}}\frac{{\cal{F}}_{b}}{q_1^2-m^2_{D^{-}}},
\end{align}
where $g_{a}=-g_{D^{*}D\gamma}g_{DN\Lambda_c}g_{\bar{T}_{cc}}$, $g_b=-ieg_{D^{*}N\Lambda_c}g_{\bar{T}_{cc}}$,
${\cal{F}}_{a}={\cal{F}}_{\bar{D}^0}{\cal{F}}_{D^{*-}}$ and ${\cal{F}}_{b}={\cal{F}}_{\bar{D}^{*0}}{\cal{F}}_{D^{-}}$.
The charge of  the hadron is in  units of $e=\sqrt{4\pi\alpha}$ with $\alpha$ being the fine-structure constant.
$\epsilon^{\mu\nu\alpha\beta}$ is the Levi-Civit\`{a} tensor with $\epsilon^{0123}=1$. Here we take $\Gamma_{D^{*},\bar{D}^{*}}=0$ MeV
for the $D^{*}$ and the $\bar{D}^{*}$ states because their widths are of the order of tens of keV.

To satisfy  gauge invariance, we introduce a generalized contact term  as
\begin{align}
{\cal{W}}^{\mu\nu}_{c}&=2g_b(-\gamma_{\mu}+{\cal{H}}q_2^{\mu})\frac{(k_1^{\nu}-p_1^{\nu})}{q^2_2-m^2_{\bar{D}^{*0}}}\frac{{\cal{F}}_{b}}{q_1^2-m^2_{D^{-}}},
\end{align}
where ${\cal{H}}=(m_p-m_{\Lambda_c^{+}})/m^2_{\bar{D}^{*0}}$ with $m_p$ and $m_{\Lambda_c^{+}}$ are the masses of proton and $\Lambda_c^{+}$, respectively.

The differential cross section in the center of mass (c.m.) frame for the $\gamma{}p\to{}D^{+}\bar{T}^{-}_{cc}\Lambda_c^{+}$
reaction reads:
\begin{align}
&d\sigma(\gamma{}p\to{}\Lambda_c^{+}D^{+}\bar{T}^{-}_{cc})\nonumber\\
&=\frac{m_p}{2(k_1\cdot{k_2})}\sum_{s_i,s_f}|-i{\cal{M}}(\gamma{}p\to{}\Lambda_c^{+}D^{+}\bar{T}^{-}_{cc})|^2\nonumber\\
&\times\frac{d^3\vec{p}_1}{2E_1}\frac{d^3\vec{p}_2}{2E_2}\frac{m_{\Lambda_c^{+}}d^3\vec{p}_2}{E_2}\delta^4(k_1+k_2-p_1-p_2-p_3),\label{cs23}
\end{align}
where $E_1$,~$E_2$ and $E_3$ stand for energies of $D^{+}$,~$\bar{T}_{cc}^-$ and final $\Lambda_c^{+}$, respectively, and
${\cal{M}}={\cal{M}}_{a}+{\cal{M}}_{b}+{\cal{M}}_{c}$ is total scattering amplitude of the
$\gamma{}p\to{}D^{+}\bar{T}^{-}_{cc}\Lambda_c^{+}$ reaction.

\section{RESULTS}\label{Sec: results}
The mass of $T^+_{cc}$ is slightly below the $D^{*}D$ threshold. It is natural to treat it as a $D^{*}D$ molecular state,
dynamically generated from the interaction of the coupled channels $D^{*+}D^{0}$ and $D^{*0}D^{+}$ in the isospin $I=0$ sector.
Considering the new structure $T^+_{cc}$ as a molecular state, its production in the $\gamma{}p\to{}D^{+}\bar{T}^{-}_{cc}\Lambda_c^{+}$
reaction is evaluated via the central diffractive mechanism  and the contact term to ensure
 gauge invariance.  However, the contributions for the $\gamma{}p\to{}D^{+}\bar{T}^{-}_{cc}\Lambda_c^{+}$
reaction from other channels, such as $s-$ and $u-$ channels, are ignored because the $s-$ and $u-$ channels involve the
creation of two additional $\bar{c}c$ quarks pair in the photon induced production and are strongly suppressed.

To make a reliable prediction for the cross-section of the $\gamma{}p\to{}D^{+}\bar{T}^{-}_{cc}\Lambda_c^{+}$ reaction, the only issue
we need to clarify is the value of $\alpha$ of the form factors.  The $\alpha$ reflects the nonperturbative
property of QCD at  low-energy scales and could not be determined from first principles.  It is usually determined from
experimental branching ratios.  We noticed  studies of $\alpha$ have been performed by comparison with  experimental
data~\cite{Belle:2007qxm, BaBar:2006qlj}, whose procedures are  illustrated in Ref.~\cite{Guo:2016iej}.  In this work, we adopt
$a=1.5$ or $1.7$ because this value is determined from the experimental data~\cite{Belle:2007qxm,BaBar:2006qlj} with the same $D$
and $D^{*}$ form factors adopted in the work of Ref.~\cite{Guo:2016iej}.

\begin{figure}[htbp]
\begin{center}
\includegraphics[bb=-30 150 800 580, clip,scale=0.33]{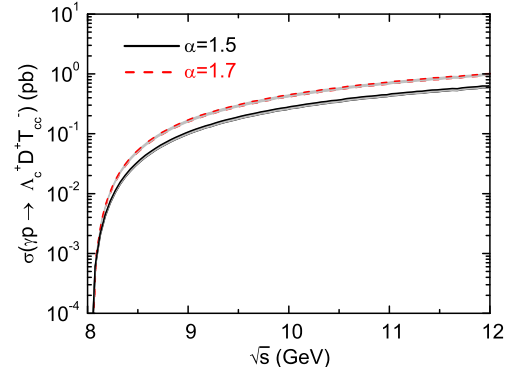}
\caption{(Color online)  Total cross section for the $\gamma{}p\to{}D^{+}\bar{T}^{-}_{cc}\Lambda_c^{+}$ reaction
with $\Gamma_{D^{*\pm}\to{}D^{\pm}\gamma}=0.979-1.704$ KeV~\cite{Zyla:2020zbs} as a function of the cutoff parameter $\alpha$.}
\label{cross-1}
\end{center}
\end{figure}
With the formalism and ingredients given above, the total cross section as a function of the C.M. energy $\sqrt{s}$ [$s=(k_1+k_2)^2$] for
the $\gamma{}p\to{}D^{+}\bar{T}^{-}_{cc}\Lambda_c^{+}$ reaction can be easily obtained.   The theoretical results obtained with a
cutoff $\alpha=1.5$ or $1.7$ for  the C.M. energy from near threshold up to 12.0 GeV are shown in Fig.~\ref{cross-1}.
The total cross section increases with $\alpha$ in the discussed cutoff range but the dependence is relatively weak.  Taking $\sqrt{s}=10$ GeV and
 $\Gamma_{D^{*\pm}\to{}D^{\pm}\gamma}=1.334$ keV as an example, and the increase of the cutoff parameter from 1.5 to 1.7 results in an increase of
 the cross section runs from 0.262 to 0.426 pb.

Fig.~\ref{cross-1} also shows that the total cross-section increases sharply near the threshold.  At higher energies,
the cross-section increases continuously but relatively slowly compared with that near threshold.  The difference between the cross-section predicted
with  $\alpha=1.5$  and that predicted with $\alpha=1.7$ becomes larger as the energy increases.
  The total cross section is about 0.7 and 0.4 pb for $\alpha=1.7$ and $\alpha=1.5$, respectively.
More concretely, for a C.M energy of about 11.0 GeV and a parameter $\alpha=1.7$ ($\alpha=1.5$) as an example, the cross-section is 0.70 (0.43) pb
for $\bar{T}^{-}_{cc}$ production,  which is accessible at EicC~\cite{Anderle:2021wcy} and US-EIC~\cite{Accardi:2012qut}
due to the higher luminosity.   It is worth noting when we increase the C.M energy to 50 GeV, which is far beyond the energy range
of the planned EIC experiment, the cross-section is also of the order of 1.0 pb.  Thus, our results suggest that it will take
at least one year of running at US-EIC to collect a hundred events.

As shown in Fig.~\ref{cross-1}, we present the variation of the total cross sections for different $\Gamma_{D^{*\pm}\to{}D^{\pm}\gamma}$ values, where
$\Gamma_{D^{*\pm}\to{}D^{\pm}\gamma}=0.979-1.704$ is taken from Ref.~\cite{Zyla:2020zbs}. The  results vary little and the
cross section can reach 0.72 pb for the $\alpha=1.7$ case at a C.M energy of $\sqrt{s}=11.0$ GeV, which should be compared with the central value of 0.7 pb.

\begin{figure}[htbp]
\begin{center}
\includegraphics[scale=0.38]{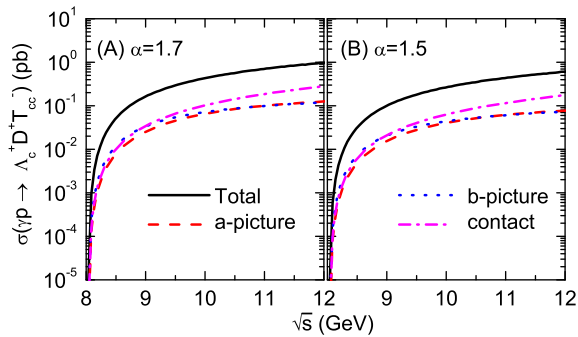}
\caption{(Color online) Individual contributions of diagram a
(red dash line), diagram b (blue dot line) and the contact term (megenta dash dot line)
for the $\gamma{}p\to{}D^{+}\bar{T}^{-}_{cc}\Lambda_c^{+}$ reaction as a function of the energy.}
\label{decaywidth}
\end{center}
\end{figure}
The individual contributions of diagrams Fig.~\ref{mku-coupling}(a), Fig.~\ref{mku-coupling}(b) and the contact term for the $\gamma{}p\to{}D^{+}\bar{T}^{-}_{cc}\Lambda_c^{+}$
reaction against the C.M energy are shown in Fig.~\ref{decaywidth}.  From the Fig the contribution from the Fig.~\ref{mku-coupling}(b) is
a little larger than all the other contributions near the threshold.  However, the contribution from the contact term becomes most important as the C.M energy $\sqrt{s}$ increase from 9.3 to 12.0 GeV.  We also notice that the
interferences among them are  sizable,  leading to a bigger total cross section.  The contribution from diagram
Fig.~\ref{mku-coupling}(a) is smaller than that from diagram  Fig.~\ref{mku-coupling}(b) near the threshold, while they become almost the same at high energies.
A possible explanation is that the low energy photon beam (still high) directly decaying to the charm meson pair $D^{+}D^{-}$ is easier than that decaying to the $D^{*-}D^{+}$ pair~\cite{R.Poling,A.Osterheld,Belle:2006hvs,Belle:2007qxm,BaBar:2006qlj}.

Strictly speaking, such a high-energy interaction should employ the Regge approach to estimate the production cross-section of $T^{+}_{cc}$ in the
$\gamma{}p\to{}D^{+}\bar{T}^{-}_{cc}\Lambda_c^{+}$ reaction, because the Regge approach is known to able to explain very well high-energy scattering with
unitarity preserved.  Moreover, a unique feature of the Regge amplitudes is that they can reproduce the diffractive pattern both at forward and
backward scatterings as well as the asymptotic behavior consistently with unitarity.  In other words, the Regge approach can be used to study the
$\gamma{}p\to{}D^{+}\bar{T}^{-}_{cc}\Lambda_c^{+}$ reaction via the current mechanism.    The relevant Regge amplitudes for the charm mesons $D$ and
$D^{*}$ exchange can be found in Ref.~\cite{Kim:2015ita}.  However, we found even the photon beam energy is very high, the energy transferred to the
$\bar{D}$ and $\bar{D}^{*}$ meson to  form the
$\bar{T}^{-}_{cc}$ state is not much, accounting for about $8-30\%$ of the total energy.  At this energy transferred, the total cross-section
is strongly suppressed from the $(s/s_0)^{\alpha_{D^{(*)}}}$ term and is of the order of $10^{-24}$ pb.  It means that the Regge approach is not
suited to study the production of $\bar{T}^{-}_{cc}$ in the $\gamma{}p\to{}D^{+}\bar{T}^{-}_{cc}\Lambda_c^{+}$ reaction.

\section{SUMMARY}
Inspired by the newly observed doubly charmed meson $T^{+}_{cc}$, we performed a detailed study of the nonresonant contribution to the
$\gamma{}p\to{}D^{+}\bar{T}^{-}_{cc}\Lambda_c^{+}$, and to estimate the $\bar{T}^{-}_{cc}$
production rate at relatively high energies,  where no data are available up to now.  The production process
is described by the central diffractive and contact term, while  the contributions for
$\gamma{}p\to{}D^{+}\bar{T}^{-}_{cc}\Lambda_c^{+}$ reaction from $s-$ and $u-$ channels are ignored  because
the $s-$ and $u-$ channels that involve the creation of two additional $\bar{c}c$ quarks pair in the photon
induced production are usually strongly suppressed.
The coupling constant of the $\bar{T}^{-}_{cc}$ to $\bar{D}^{*}\bar{D}$ are obtained from chiral unitary
theory~\cite{Feijoo:2021ppq}, where the $\bar{T}^{-}_{cc}$ is dynamically generated.

Our study showed that the cross section for the $\gamma{}p\to{}D^{+}\bar{T}^{-}_{cc}\Lambda_c^{+}$ reaction can reach 1.0 pb.  Although the photoproduction cross section is quite small, it is still possible to
test our theoretical predictions thanks to the low background of the exclusive and specific reaction proposed in this work. The future
electron-ion colliders (EIC) of high luminosity in US ($10^{34}$ cm$^{-2}$s$^{-1}$) \cite{AbdulKhalek:2021gbh,Accardi:2012qut}
and China ($2-4\times10^{33}$ cm$^{-2}$s$^{-1}$)\cite{ Chen:2018wyz,Chen:2020ijn,Anderle:2021wcy} provide a good platform for this purpose. For a
year of running, around a hundred events can be collected.  In experiment, it is vital to improve the detection efficiency of the multiple
final-state particles of this reaction. To take advantage of the virtual photon of small
virtuality on EIC, the far-forward electron detector is needed. The Roman pot frequently discussed could be used for this purpose.
The photoproduction channel should be explored to study the exotic state $T^{+}_{cc}$.

\section*{Acknowledgments}
This work was supported by the Science and Technology
Research Program of Chongqing Municipal Education Commission (Grant No. KJQN201800510), the Opened Fund
of the State Key Laboratory on Integrated Optoelectronics
(GrantNo. IOSKL2017KF19). Yin Huang want to thanks
the support from the Development and Exchange Platform for
the Theoretic Physics of Southwest Jiaotong University under Grants No.11947404 and No.12047576,
the Fundamental Research Funds for the Central Universities(Grant No.
2682020CX70), and the National Natural Science Foundation
of China under Grant No.12005177.

\end{document}